\documentstyle[12pt,epsfig]{article}
\begin{document}

\newcommand{\bea}{\begin{eqnarray}}
\newcommand{\eea}{\end{eqnarray}}
\newcommand{\beq}{\begin{equation}}
\newcommand{\eeq}{\end{equation}}
\newcommand{\bay}{\begin{array}}
\newcommand{\eay}{\end{array}}

\rightline{CLNS-99/1604}
\rightline{TECHNION-PH-99-7}
\bigskip
\bigskip
\centerline{\bf A Critical Look at Rescattering Effects on $\gamma$ from 
$B^+\to K\pi$}
\bigskip\bigskip\bigskip
\centerline{Michael Gronau}
\medskip
\centerline{\it Department of Physics}
\centerline{\it Technion -- Israel Institute of Technology, Haifa 32000,
Israel}
\medskip
\centerline{and}
\medskip
\centerline{Dan Pirjol}
\medskip
\centerline{\it Floyd R. Newman Laboratory of Nuclear Studies} 
\centerline{\it Cornell University, Ithaca, New York 14853}
\bigskip\bigskip\bigskip
\centerline{\bf ABSTRACT}
\medskip
\begin{quote}
Three ways are compared of dealing with rescattering effects in 
$B^{\pm}\to K^0\pi^{\pm}$, in order to determine the weak phase $\gamma$ 
from these processes and $B^{\pm}\to K^{\pm}\pi^0$. We find that 
neglecting these contributions altogether may involve sizable
errors in $\gamma$, depending on the rescattering amplitude and on the 
value of 
a certain measurable strong phase. 
We show that an attempt to eliminate these effects by using the 
charge-averaged rate of $B^{\pm}\to K^{\pm}K^0$ suffers from a large 
theoretical
error due to SU(3) breaking, which may be resolved when using also the 
processes 
$B^{\pm}\to \pi^{\pm}\eta_8$.
\end{quote}
\medskip
\leftline{\qquad PACS codes:  12.15.Hh, 12.15.Ji, 13.25.Hw, 14.40.Nd}
\newpage

\section{Introduction}

The weak phase $\gamma=$ Arg$(V_{ub}^*)$ is presently the least well known 
quantity among the four parameters (three angles and a phase) of the 
Cabibbo-Kobayashi-Maskawa (CKM) matrix. 
Its determination, which is regarded to be more difficult than that of the 
other two
angles of the CKM unitarity triangle \cite{Review,FLreview}, can provide a crucial 
test of the 
CKM mechanism for CP violation in the Standard Model.
Several methods have been proposed to determine $\gamma$ from hadronic 
two-body 
$B$ decays. The methods which seem to be experimentally most 
feasible in the near future are based on applications of SU(3) flavor 
symmetry in 
$B$ decays into two light charmless pseudoscalars \cite{GHLR}. 
These methods involve certain theoretical uncertainties, which are expected to
be reduced when more data become available and when better theoretical
understanding of hadronic $B$ decays is achieved.

In a first paper in a series, Gronau, London and Rosner (GLR) \cite{GRL} 
proposed to extract $\gamma$ by combining decay rate measurements of 
$B^+\to K\pi$, $B^+\to\pi\pi$ with their charge-conjugates. SU(3) breaking, 
occuring in a relation between $B\to \pi\pi~I=2$ and $B\to K\pi ~I=3/2$ 
amplitudes, was introduced through a factor $f_K/f_{\pi}$ when assuming that 
these amplitudes factorize.
In its original version, suggested before the observation of the heavy top 
quark, the method of Ref.~\cite{GRL} neglected electroweak
penguin (EWP) contributions and certain rescattering effects. Subsequently,
model-calculations showed that due to the heavy top quark the neglected EWP 
terms were significant
\cite{EWP}; and recently these terms were related by SU(3) to the $B\to K\pi~
I=3/2$ current-current amplitudes \cite{NR1,GPY}. This led to a modification 
\cite{NR2} of the GLR method, to be referred to as the GLRN method, which in 
the limit of flavor SU(3) symmetry includes EWP effects in a model-independent 
way. Corrections from SU(3) breaking, affecting the relation between EWP
terms and current-current terms, were argued to be small \cite{NR1,Neubert}. 

Assuming that the above SU(3) breaking effects are indeed under control, there 
is still an uncertainty due to rescattering effects. To determine 
$\gamma$ from the above rates, one takes the $B^+\to K^0\pi^+$ amplitude to be 
pure penguin, involving no term with weak phase $\gamma$. This assumption, which
neglects quark annihilation and rescattering contributions from charmless 
intermediate states, was challenged by a large number of authors \cite{Rescat}.
Several authors proposed ways of controlling rescattering effects in 
$B^{\pm}\to K^0\pi^{\pm}$ by relating them through SU(3) to the much enhanced 
effects in $B^{\pm}\to K^{\pm}\bar K^0$ \cite{Falk,GR,FL} (see also 
\cite{He,DeshAg,AtSo}). 
The  charge-averaged rate of the latter processes can be used to set an upper 
limit on the rescattering amplitude in $B^{\pm}\to K^0\pi^{\pm}$. While present 
limits are at the level of $20-30\%$ of the dominant penguin amplitude 
\cite{GPY,Neubert} (depending somewhat on the value of $\gamma$), they are 
expected to be improved in the future. 
The smaller the rescattering amplitude is, the more precisely can $\gamma$ be 
determined from the GLRN method. A recent demonstration \cite{Neubert}, based 
on a few possible rate measurements, seems to show that if the rescattering 
amplitude is an order of magnitude smaller than the dominant penguin amplitude 
in $B^+\to K^0\pi^+$, the uncertainty in $\gamma$ is only about 5 degrees. 

In the present Letter we reexamine in detail the uncertainty in $\gamma$ 
due to rescattering effects. Using a geometrical interpretation for the 
extraction of $\gamma$, we perform in Section 2 numerical simulations which 
cover the entire 
parameter space of the two relevant strong phases, the rescattering phase 
$\phi_A$ and the relative phase $\phi$ between $I=3/2$ current-current and 
penguin amplitudes. We find that, contrary to the demonstration made in 
\cite{Neubert}, a 10$\%$ rescattering amplitude leads to an uncertainty 
in $\gamma$ as large 
as about $14^{\circ}$ around $\phi\sim 90^{\circ}$. For certain singular cases 
no solution can be found for $\gamma$. We show that $\phi$ can 
be determined rather precisely from the $B^{\pm}\to K\pi$ rate measurements 
\cite{Neubert}, which could reduce substantially the error in $\gamma$ if values
far apart from $\phi=90^{\circ}$ were found. 

It has been suggested \cite{FL} to go one step beyond setting limits 
on rescattering contributions in $A(B^\pm\to K^0\pi^\pm)$ and to completely 
eliminate them by using the charge-averaged rate measurement of 
$B^{\pm}\to K^{\pm}K^0$. Applying our geometrical formulation, we will show in 
Section 3 that the resulting determination of $\gamma$ is unstable under SU(3) 
breaking which can introduce very large uncertainties in $\gamma$ . 

Finally, in order to overcome these uncertainties, we have recently proposed to
use in addition to $B^{\pm}\to K^{\pm} \bar K^0$ also the processes 
$B^{\pm} \to \pi^{\pm} \eta_8$ \cite{GP}. Although this may be considered an
academic exercise, mainly due to complicating $\eta-\eta'$ mixing effects, 
we will examine in Section 4 the precision of this method. We will show that, 
when neglecting $\eta-\eta'$ mixing, the theoretical error in $\gamma$ is 
reduced to a few degrees. We conclude in Section 5. An algebraic condition, 
used in Section 3 to eliminate rescattering effects by 
$B^\pm\to K^{\pm}K^0$ decays, is derived in an Appendix.

\section{Rescattering uncertainty in $\gamma$ from $B^{\pm}\to K\pi$}

The amplitudes for charged $B$ decays can be parameterized in terms of graphical 
contributions representing SU(3) amplitudes (we use the notations of \cite{GPY}):
\bea\label{B1}
A(B^+\to K^0\pi^+) &=& |\lambda_u^{(s)}|e^{i\gamma} (A+P_{uc}) +
                       \lambda_t^{(s)} (P_{ct} + P_3^{EW})~,\\
\label{B2}
\sqrt2 A(B^+\to K^+\pi^0) &=& |\lambda_u^{(s)}|e^{i\gamma} (-T-C-A-P_{uc}) +
                       \lambda_t^{(s)} (-P_{ct} + \sqrt2 P_4^{EW})~,\\
\label{B2'}
\sqrt2 A(B^+\to \pi^+\pi^0) &=& |\lambda_u^{(s)}|e^{i\gamma} (-T-C)~,
\eea
where $\lambda_{q'}^{(q)}=V^*_{q'b}V_{q'q}$ are the corresponding CKM factors.
These amplitudes satisfy a triangle relation 
\cite{GRL,NR2}
\bea\nonumber
& &\sqrt2 A(B^+\to K^+\pi^0) + A(B^+\to K^0\pi^+) =
\sqrt2 \tilde r_u
|A(B^+\to\pi^+\pi^0)| e^{i(\gamma+\xi)} 
\left(1 - \delta_{EW} e^{-i\gamma}\right)~.\\\label{4}
\eea
Here we denote $\tilde r_u = (f_K/f_{\pi})\lambda/(1-\lambda^2/2)\simeq 0.28,~  
\delta_{EW}=-(3/2)|\lambda^{(s)}_t/\lambda^{(s)}_u|\kappa \simeq 0.66$ 
($\kappa\equiv (c_9+c_{10})/(c_1+c_2)=-8.8\cdot 10^{-3}$), while $\xi$ 
is an unknown strong phase. The second term in the brackets represents the sum 
of EWP contributions to the amplitudes on the left-hand side \cite{NR1,GPY}.
The factor $f_K/f_\pi$ accounts for factorizable SU(3) breaking 
effects.

The relation (\ref{4}), together with its charge-conjugate counterpart, written 
for $\tilde A(\bar B\to\bar f)\equiv e^{2i\gamma} A(\bar B\to\bar f)$,
are represented graphically by the two triangles OAA$'$ and OBB$'$ in Fig.~1. 
Here all amplitudes are divided by a common factor ${\cal A}\equiv \sqrt2
\tilde r_u |A(B^+\to\pi^+\pi^0)|e^{i(\gamma+\xi)}$, such that the horizontal line 
$OI$ is of unit
length and the radius of the circle is $\delta_{EW}$. Four of the sides of the 
two triangles are given by
\bea\label{x}
x_{0+} &=& \frac{1}{\sqrt 2\tilde r_u}\frac{|A(B^+\to K^0\pi^+)|}
{|A(B^+\to\pi^+\pi^0)|}~,\qquad
x_{+0} = \frac{1}{\tilde r_u}\frac{|A(B^+\to K^+\pi^0)|}
{|A(B^+\to\pi^+\pi^0)|}~,\\
\nonumber
\tilde x_{0-} &=& \frac{1}{\sqrt 2\tilde r_u}
\frac{|A(\bar B^-\to \bar K^0\pi^-)|}{|A(B^+\to\pi^+\pi^0)|}~,\qquad
\tilde x_{-0} = \frac{1}{\tilde r_u}
\frac{|A(\bar B^-\to K^-\pi^0)|}{|A(B^+\to\pi^+\pi^0)|}~.
\eea
The relative orientation of the two triangles depends on $\gamma$ and is not 
determined from measurements of the sides alone. {\it Assuming that the 
rescattering amplitude with weak phase $\gamma$ in $B^+\to K^0\pi^+$ can be 
neglected}, one 
takes the amplitude (\ref{B1}) to be given approximately by the second 
(penguin) term \cite{GRL,NR2}, which implies $OB=e^{2i\gamma}OA$ in Fig. 1. 
In this approximation, the weak phase $\gamma$ is determined by requiring that 
the angle ($2\gamma$) between $OA$ and $OB$ is equal to the angle ($2\gamma$) 
at the center of the circle \cite{NR2}. 

In order to study the precision of determining in this way the phase 
$\gamma$ as function of the rescattering contribution which is being neglected, 
let us rewrite (\ref{B1}) in the form
\bea\label{epsA}
A(B^+\to K^0\pi^+) = -V_{cb}\left(1-\frac{\lambda^2}{2}\right)p
(1 + \epsilon_A e^{i\phi_A}e^{i\gamma})~,~~~~~p\equiv P_{ct}+P_3^{EW}
\eea
where $\epsilon_A$ measures the magnitude of rescattering effects.
In Fig.~1 the magnitude of these effects has a simple geometrical 
interpretation in terms of the distance of the point $Y$ from the origin 
$O,~~\epsilon_A=|YO|/|YA|$, where $YO$ and $YA$ are the two components in the 
$B^+\to K^0\pi^+$ amplitude carrying weak phases $\gamma$ and zero, 
respectively
\bea\label{OY}
YO = |\lambda_u^{(s)}|e^{i\gamma}[(A+P_{uc}) - p]/{\cal A}~,~~~~~~
YA = V_{cb}\left(1-\frac{\lambda^2}{2}\right)p/{\cal A}~.
\eea
The rescattering phase $\phi_A$ is given by $\phi_A=$Arg$(YO/YZ)$, where $Z$ is
any point on the line bisecting the angle $AYB$. A second strong phase 
which affects the determination of $\gamma$ 
is $\phi$, the relative strong phase 
between the penguin amplitude $p$ and the $I=3/2$ current-current amplitude
$T+C$. In Fig.~1 this phase is given by $\phi=$Arg$(YZ/OI)$. 

Let us now investigate the dependence of the error in $\gamma$ when
neglecting rescattering on the relevant hadronic parameters.
Our procedure will be as follows. First we generate a set of amplitudes 
based on the geometry of Fig.~1 and on given values of the  
parameters $\gamma,~\epsilon,~\epsilon_A,\phi_A$ and $\phi$; then we
solve the equation $\cos 2\gamma=\cos(BOA)$ and compare the output value
of $\gamma$ with its input value. Here $\epsilon$ is given in terms of the 
ratio of charge-averaged branching ratios \cite{GRL,NR2}
\beq
\epsilon\equiv \frac{\lambda}{1-\lambda^2/2} \frac{f_K}{f_\pi}
\sqrt{\frac{2B(B^\pm\to\pi^\pm\pi^0)}{B(B^\pm\to K^0\pi^\pm)}}~,
\eeq
The geometrical construction in Fig.~1 is described by 
\bea
YA = \frac{e^{i(\phi-\gamma)}}{\epsilon\sqrt{1+2\epsilon_A\cos\phi_A\cos\gamma+
\epsilon_A^2}}
OI~,\qquad
OY = \epsilon_A e^{i(\phi_A+\gamma)}YA~,
\eea
implying a rate asymmetry between $B^+\to K^0\pi^+$ and $B^-\to \bar K^0\pi^-$.

For illustration, we take $\gamma=76^{\circ}, \epsilon=0.24$ \cite{NR1}, 
$\epsilon_A=0.1$ (which is a reasonable guess \cite{Falk,GR}), and we vary 
$\phi$ and $\phi_A$ in the range $0^{\circ}\le \phi \le 180^{\circ},~
-90^{\circ}\le \phi_A \le 270^{\circ}$.
The results of a search for solutions in the interval $65^\circ\le\gamma\le
90^\circ$ are presented in Fig.~2 which displays a twofold ambiguity.  
Fig. 2(a) shows the solution as function of $\phi_A$ for two values of $\phi$, 
$\phi=60^{\circ}$ and $\phi=90^{\circ}$. Whereas for $\phi_A=90^{\circ}$ the 
solution is very close to the input value, the deviation becomes maximal for
$\phi_A=0^{\circ},180^{\circ}$. This agrees with the geometry of Fig.~1, in 
which the largest rescattering effects are expected when $YO$ is parallel or
anti-parallel to the line bisecting the angle $BYA$.

In a second plot, Fig.~2(b), we fix $\phi_A=0^\circ$ and vary $\phi$ over its
entire range, which illustrates the maximal rescattering effect.
We find two branches of the solution for $\gamma$, both of which deviate 
strongly from the input value $\gamma=76^\circ$ for values of $\phi$ around 
$90^\circ$. At $\phi=90^{\circ}$ there is no solution for $\epsilon_A=0.1$ 
in the considered interval. We checked that the solution is restored and 
approaches the input value as the magnitude of $\epsilon_A$ decreases to zero, 
as it should. Thus, the uncertainty in $\gamma$, seen both in Fig.~2(a) and 
Fig.~2(b) at $\phi_A=0^{\circ}$ and around $\phi=90^{\circ}$, is about 
$14^{\circ}$. It can even be worse in the singular cases where no solution for
$\gamma$ can be found.

A variant of this method for determining $\gamma$, proposed recently in 
\cite{Neubert}, was formulated in terms of two quantities $R_*$ and $\tilde A$ 
defined by
\bea
& &R_* \equiv \frac{B(B^\pm\to K^0\pi^\pm)}{2B(B^\pm\to K^\pm\pi^0)}~,
\\
& &\tilde A \equiv \frac{B(B^+\to K^+\pi^0)-B(B^-\to K^-\pi^0)}{B(B^\pm\to 
K^0\pi^\pm)} - \frac{B(B^+\to K^0\pi^+)-B(B^-\to \bar K^0\pi^-)}{2B(B^\pm\to 
K^0\pi^\pm)}~.\nonumber
\eea
These quantities do not contain ${\cal O}(\epsilon_A)$ terms; their dependence 
on the rescattering parameter $\epsilon_A$ appears only at order 
${\cal O}(\epsilon\epsilon_A)$. Therefore, it was argued in \cite{Neubert}, 
the determination of $\gamma$, by setting $\epsilon_A=0$ in the expressions for 
$R_*$ and $\tilde A$, is insensitive to rescattering effects. 
This procedure gives two equations for $\gamma$ and $\phi$ which can 
be solved simultaneously from $R_*$ and $\tilde A$. Using two pairs of input 
values for ($R_*, \tilde A$) (corresponding to a restricted range for $\phi_A$ 
and $\phi$) seemed to indicate that the error in $\gamma$
for $\epsilon_A=0.08$ is only about $5^{\circ}$. (the relations between the
parameters used in \cite{Neubert} and ours are $\phi=-\phi, \eta=\phi_A+\pi,
\bar\epsilon_{3/2}=\epsilon$ and $\epsilon_a=\epsilon_A$). 

In Fig.~3 we show the results of such an analysis carried out for the entire 
parameter space of $\phi_A$ and $\phi$.  
Whereas the angle $\phi$ can
be recovered with small errors, the results for $\gamma$ show the same 
large rescattering effects for values of $\phi$ around 90$^\circ$ as in Fig.~2.
(A slight improvement is the absence of a discrete ambiguity in the value of 
$\gamma$.) These results show that the large deviation of $\gamma$ from its 
physical value for $\phi=90^\circ$ is a general phenomenon, common to all 
variants of this methods. Some information about the size of the
expected error can be obtained by first determining $\phi$. Values not too 
close to 90$^\circ$ would be an indication for a small error. 

\section{Eliminating rescattering by $B^{\pm}\to K^{\pm}K^0$}

The amplitude for $B^+\to K^+\bar K^0$ is obtained from $A(B^+\to K^0 \pi^+)$
in (\ref{B1}) by a $U$-spin rotation \cite{Falk}
\bea\label{K+K}
A(B^+\to K^+\bar K^0) &=& |\lambda_u^{(d)}|e^{i\gamma} (A+P_{uc}) +
|\lambda_t^{(d)}|e^{-i\beta}(P_{ct} + P_3^{EW})~.
\eea
In the limit of SU(3) symmetry the amplitudes in (\ref{K+K}) are exactly
the same as those appearing in (\ref{B1}). In Fig.~4 $A(B^+\to K^+\bar K^0)$, 
scaled by the factor $\lambda/(1-\lambda^2/2)$ (and divided by ${\cal A}$ as in 
Fig.~1), is given by the line $OC$ and its charge-conjugate is given by $OD$.
We have shown in \cite{GP} that knowledge of these two amplitudes allows one
to completely eliminate the rescattering contribution $A+P_{uc}$ from the 
determination of $\gamma$. This is achieved by effectively replacing in the 
GLRN method the origin $O$ by the intersection $Y$ of the lines $AC$ and $BD$.
$\gamma$ is determined by requiring that the angle ($2\gamma$) between $YA$ and 
$YB$ is equal to the angle ($2\gamma$) at the center of the circle. 

The amplitude (\ref{K+K}) can be decomposed into two terms carrying definite 
weak phases in form very similar to (\ref{epsA}), 
\bea\label{epsAK}
\frac{\lambda}{1-\lambda^2/2}A(B^+\to K^+\bar K^0) = 
-V_{cb}\left(1-\frac{\lambda^2}{2}\right)p
\left(-\frac{\lambda^2}{(1-\lambda^2/2)^2} + \epsilon_A e^{i\phi_A}e^{i\gamma}
\right)~,
\eea
The ratio $|CY|/|AY| = \lambda^2/(1-\lambda^2/2)^2$ implies that 
the triangle $AYB$ is about 25 times larger than the triangle $CYD$.
This will result in a large uncertainty in $\gamma$ also when the equality 
between the corresponding terms in $B^+\to K^0\pi^+$ and $B^+\to K^+\bar K^0$ 
amplitudes involves relatively small SU(3) violation.

The geometrical construction by which rescattering amplitudes can be completely 
eliminated in the SU(3) limit consists of three steps. (See Fig.~4. For an 
alternative suggestion, see \cite{FL}.)

a) Determine the position of the point $Y$ as a function of the variable angle 
$2\gamma$ and the decay rates of $B^\pm\to K\pi$ and $B^+\to \pi^+\pi^0$. The 
point $Y$ is chosen on the mid-perpendicular of $AB$ such that the equality of 
the angles marked $2\gamma$ is preserved for any value of $\gamma$.

b) Draw two circles of radii $\lambda/(1-\lambda^2/2)|A(B^\pm\to K^0 K^\pm)|$ 
centered at the origin $O$ (dashed-dotted circles in Fig.~4).
The intersections of the lines $AY$ and $BY$ with these circles determine $C$ 
and $D$ respectively (up to a two-fold ambiguity), again as functions of 
$\gamma$. 

c) The physical value of $\gamma$ is determined by the requirement $|AC|=|BD|$
\cite{GP}.
\noindent
This condition on $\gamma$ can be formulated in an algebraic form, showing
that only the charge-averaged rate of $B^{\pm}\to K^{\pm}K^0$ is needed. The
condition is given by Eq.~(\ref{cond}) in the Appendix.

Let us examine the precision of this method for $\epsilon_A=0.1$
at $\phi\simeq 90^\circ$, for 
which the simpler method of Sec.~2 receives large rescattering corrections. 
In Fig.~5(a) we show the left-hand side of Eq.~(\ref{cond})
as a function of variable $\gamma$ at $\phi= 90^\circ$ for several values of 
$\phi_A$. The value of $\gamma$ 
%($\gamma=76^{\circ}$) 
is obtained from the condition that the left-hand side of this equation vanishes.
In the absence of SU(3) breaking this method reproduces precisely the physical 
value of $\gamma$ ($\gamma=76^{\circ}$) for all values of $\phi_A$. However SU(3) 
breaking effects can
become important, to the point of completely spoiling this method. We simulate
these effects by taking the amplitudes $p$ and $a\equiv A+P_{uc}-p$ in $B^\pm\to 
K^\pm K^0$ (Eq.~(\ref{K+K})) to differ by at most 30\% from those 
in $B^\pm\to K^0 \pi^\pm$ (Eq.~(\ref{B1})). 
This expands the lines of Fig.~5(a) into bands of finite width, which give
a range for the output value of $\gamma$.

In Fig.~5(b) we show the effects of SU(3) breaking on the determination of 
$\gamma$ as function of $\phi_A$ for $\phi = 90^\circ$. We see that for values
of $|\phi_A|$ larger than about 25$^\circ$ the error on $\gamma$ is quite
large.
%left-hand side of (\ref{cond}) vanishes for values of $\gamma$ which are quite 
%larger than the input value. 
Thus, we conclude that for certain values of the
strong phases the determination of $\gamma$ using this method
is unstable under SU(3) breaking in 
the relation between $B^+\to K^0\pi^+$ and $B^+\to K^+ \bar K^0$. 

\section{The use of $B^{\pm}\to \pi^{\pm}\eta_8$}

In Ref.~\cite{GP} we proposed to use in addition to $B^+\to K^+ \bar K^0$ also
$B^+\to \pi^+\eta_8$ and their charge-conjugates. Writing
\beq
A(B^+\to \pi^+\eta_8)=|\lambda^{(d)}_u|e^{i\gamma}(-T-C-2A-2P_{uc}) +
|\lambda^{(d)}_t|e^{-i\beta}(-P_{ct}+P_5^{EW})~,
\eeq
we find the triangle relation
\beq
A(B^+\to K^+\bar K^0) + \sqrt{\frac{3}{2}}A(B^+\to\pi^+\eta_8)=\frac{1}{\sqrt 2}
A(B^+\to \pi^+\pi^0)~.
\eeq
This relation and its charge-conjugate provide another condition which 
determines the positions of the points $C$ and $D$. As in Section 3, the phase 
$\gamma$ is determined by the equation $\cos(BYA) = \cos 2\gamma$, where the 
point $Y$ is fixed by the intersection of the lines $AC$ and $BD$.
General considerations, based on the relative sizes of the amplitudes
involved, suggest that this method is relatively insensitive to SU(3) breaking
effects \cite{GP}.

We illustrate this in Fig.~6 where we show on the same plot the two sides
of the equation $\cos(BYA) = \cos 2\gamma$ as functions of the variable $\gamma$.
As in method of Section 3, SU(3) breaking is simulated by taking the penguin 
($p$) and annihilation ($a$) amplitudes in $B^\pm\to K^\pm K^0$ to differ by at 
most 30\% (separately for their real and imaginary parts) from those in 
$B^\pm\to K^0 \pi^\pm$. The latter are used to construct the positions of the 
points $C$ and $D$.
In the example of Fig.~6 we take $\epsilon_A=0.1, \phi=90^\circ, \phi_A=
45^\circ$, for which the two methods described
in Sections 2 and 3 were shown to lead to large errors in $\gamma$.
For an input value $\gamma=76^{\circ}$, the output is given by the range
$74^{\circ}<\gamma<78^{\circ}$, obtained by the
intersection of the solid line with the band formed by the diamond points. 
We see that the error in $\gamma$ due to SU(3) breaking is less than $\pm
2^\circ$, which confirms the general arguments of \cite{GP}. 
This scheme, or rather its analogous version using $B^0$ and $B_s$ decay 
\cite{GP}, may prove useful for a determination of $\gamma$ in case that the 
strong phases $(\phi,\phi_A)$ turn out to have values which preclude the use of 
the two simpler methods.

\section{Conclusion}

We compared three ways of dealing with rescattering effects in $B^{\pm}\to K^0
\pi^{\pm}$, in order to achieve a precise determination for the weak phase 
$\gamma$ from these processes and $B^{\pm}\to K^{\pm}\pi^0$. In the simplest 
GLRN method we find that large errors in $\gamma$ are possible for a particular
range of the strong phases, $\phi\sim 90^{\circ}$, even when the rescattering 
term is only at a level of 10$\%$.~~$B^+$ and $B^-$ decay rate measurements 
are expected to provide rather precise information on $\phi$. Small errors in 
$\gamma$ would be implied if $\phi$ turns out to be far away from $90^{\circ}$.
The second method, in which
rescattering effects can be completely eliminated in the SU(3) limit by using 
also the charge-averaged $B^{\pm}\to K^{\pm}K^0$ rate, suffers from a sizable 
uncertainty due to SU(3) breaking. These uncertainties would be
resolved in an ideal world, where $B^{\pm}\to \pi^{\pm}\eta_8$ can be measured, 
or alternatively by using corresponding $B^0$ and $B_s$ decays.

\section{Appendix}

The weak angle $\gamma$ is fixed in the method described in Sec.~3 by the condition 
$|AC|=|BD|$, or equivalently $|YC|=|YD|$. Explicitly, this can be written after some 
algebra as an equation in $\gamma$ 
\bea\label{cond}
2(1-x_0)^2\vec Y^2 + 2x_0(1-x_0)\vec Y\cdot(\vec A+\vec B) + x_0^2(x_{0+}^2 + 
\tilde x_{0-}^2)- (y_{+0}^2 + \tilde y_{-0}^2) = 0~,
\eea
where $x_0$ is defined as the ratio of two CP rate differences
\beq
x_0 \equiv (y_{+0}^2 - \tilde y_{-0}^2)/(x_{0+}^2 - \tilde x_{0-}^2)
\stackrel{\longrightarrow}{\mbox{\tiny SU(3)}} -\lambda^2/(1-\lambda^2/2)^2~.
\eeq
Here 
\bea
y_{+0} &=& \frac{1}{\sqrt 2}\frac{f_{\pi}}{f_K} 
\frac{|A(\bar B^+\to K^+\bar K^0)|}{|A(B^+\to\pi^+\pi^0)|}~,\qquad
\tilde y_{-0} = \frac{1}{\sqrt 2}\frac{f_{\pi}}{f_K}
\frac{|A(\bar B^-\to K^- K^0)|}{|A(B^+\to\pi^+\pi^0)|}~
\eea
obey an SU(3) relation with the amplitudes (\ref{x}) of $B^{\pm}\to K\pi^{\pm}$
\bea\label{SU3}
y_{+0}^2 - \tilde y_{-0}^2 = -\frac{\lambda^2}{(1-\lambda^2/2)^2}(x_{0+}^2-
\tilde x_{0-}^2)~.
\eea
This implies that CP rate differences in $B^\pm\to K^0\pi^\pm$ 
and $B^\pm\to K^\pm K^0$ are equal and of opposite sign \cite{FL}.
We see that in the SU(3) limit the condition (\ref{cond}), which eliminates
rescattering effects, requires only a measurement of the charge-averaged rate of
$B^{\pm}\to K^{\pm}K^0$ and not the CP asymmetry in these processes \cite{FL}.

To prove (\ref{cond}), let us consider two lines $AY$ and $BY$ cutting two circles
of radii $R_1$, $R_2$ (centered at the origin) at points $C$ and $D$ 
respectively.
The intersection points can be written as $\vec C=\vec 
Y+x_1(\vec A-\vec Y)$ and
$\vec D=\vec Y+x_2(\vec B-\vec Y)$, where $x_1,x_2$ are solutions of the 
equations
\bea\label{A1}
& &(\vec A-\vec Y)^2 x_1^2 + 2x_1\vec Y\cdot(\vec A-\vec Y) + (\vec Y^2-R_1^2)=0\\
\label{A2}
& &(\vec B-\vec Y)^2 x_2^2 + 2x_2\vec Y\cdot(\vec B-\vec Y) + (\vec Y^2-R_2^2)=0\,.
\eea
The condition $|YC|=|YD|$ is equivalent to requiring that these two equations 
have
a common solution $x_1=x_2$. Obviously, if such a solution exists, it is given by
\bea\label{A3}
x_0 = \frac{R_1^2-R_2^2}{2\vec Y\cdot(\vec A-\vec B)} =
\frac{R_1^2-R_2^2}{\vec A^2-\vec B^2}\,,
\eea
where we used the equality $(\vec A-\vec Y)^2=(\vec B-\vec Y)^2$. 
Taking the sum of (\ref{A1}) and (\ref{A2}) with the value (\ref{A3}) for $x$
leads immediately to the condition (\ref{cond}). \vspace*{5mm}

{\em Acknowledgements.}
We thank R. Fleischer, M. Neubert and T.M. Yan for useful discussions.
This work is supported by the National Science 
Foundation and by the United States - Israel Binational Science Foundation 
under Research Grant Agreement 94-00253/3.

\newpage
\thispagestyle{plain}

\begin{figure}[hhh]
 \begin{center}
 \mbox{\epsfig{file=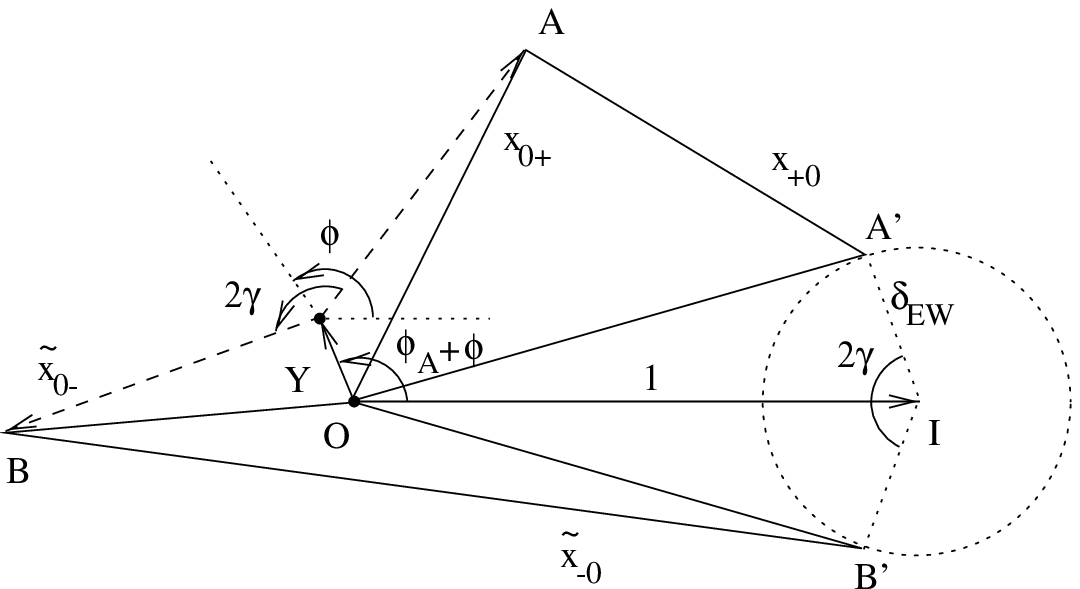,width=10cm}}
% \mbox{\epsfig{file=gamma3.eps,width=7cm}}
 \end{center}
 \caption{Relative orientation of $B^+\to K\pi$ amplitude triangles. 
}
\label{fig1}
\end{figure}

\begin{figure}[hhh]
 \begin{center}
 \mbox{\epsfig{file=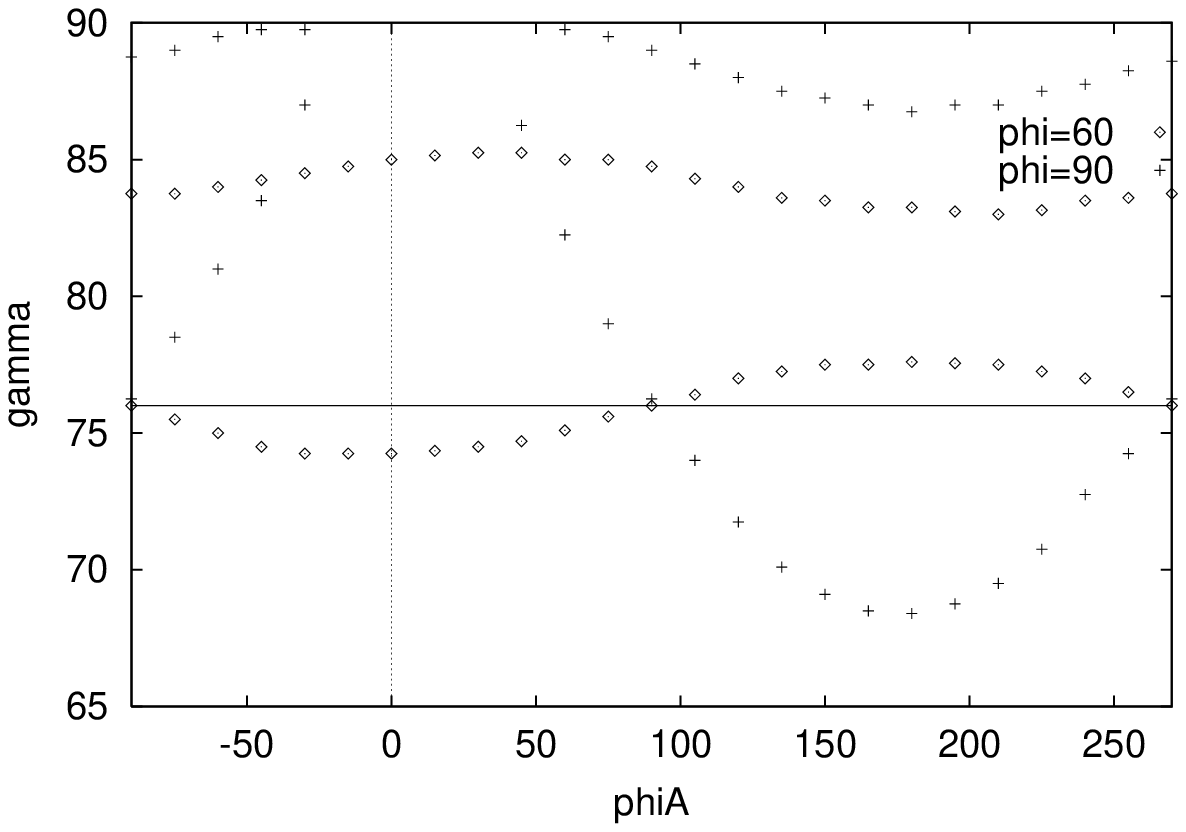,width=7cm}}
 \mbox{\epsfig{file=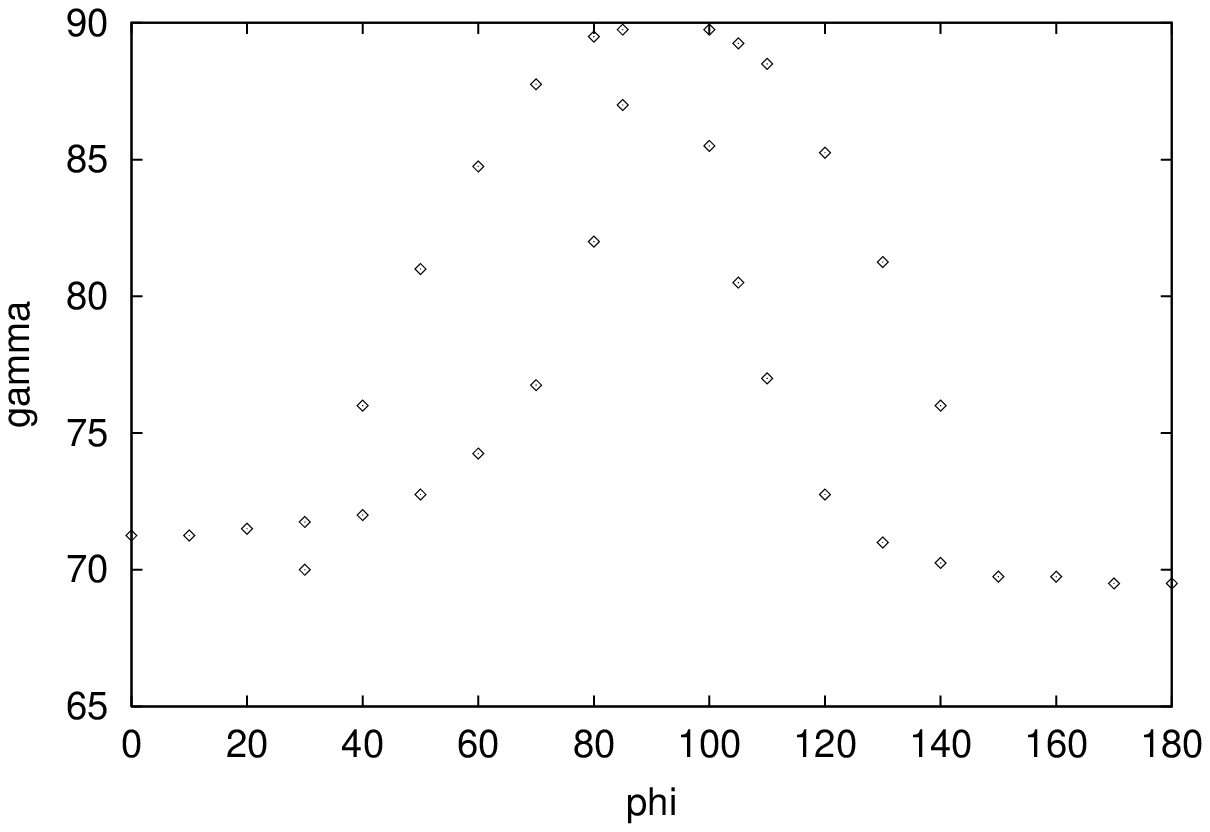,width=7cm}}\\
\hspace*{5mm}\mbox{(a)}\hspace*{7cm}\mbox{(b)}
 \end{center}
 \caption{The weak phase $\gamma$ is obtained as the solution to the equation
$\cos(2\gamma)=\cos(BOA)$. (a) - the dependence of the solution on $\phi_A$,
for two values of $\phi=60^\circ$ and $\phi=90^\circ$; (b) - the dependence
of the solution on $\phi$, for $\phi_A=0^\circ$. (both graphs correspond to
$\epsilon_A=0.1, \gamma=76^\circ$) }
\label{fig2}
\end{figure}

\begin{figure}[hhh]
 \begin{center}
 \mbox{\epsfig{file=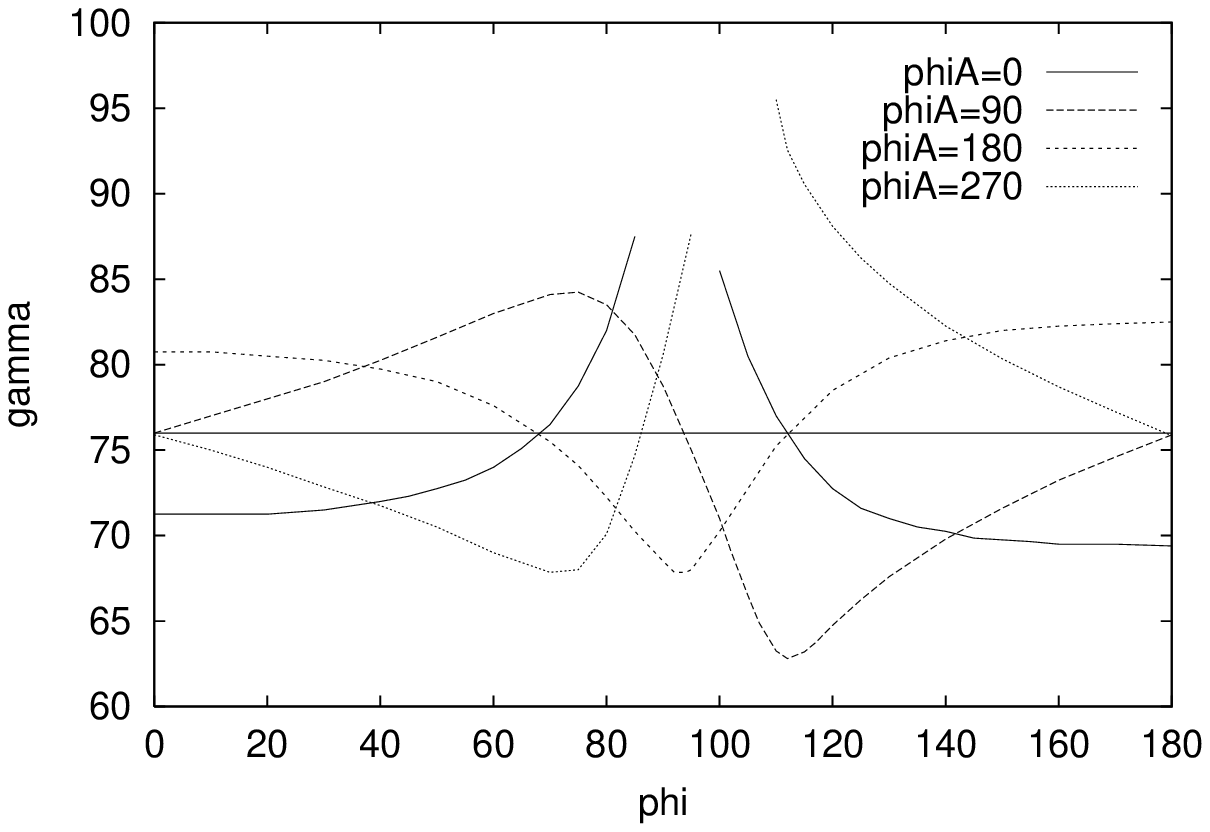,width=7cm}}
 \mbox{\epsfig{file=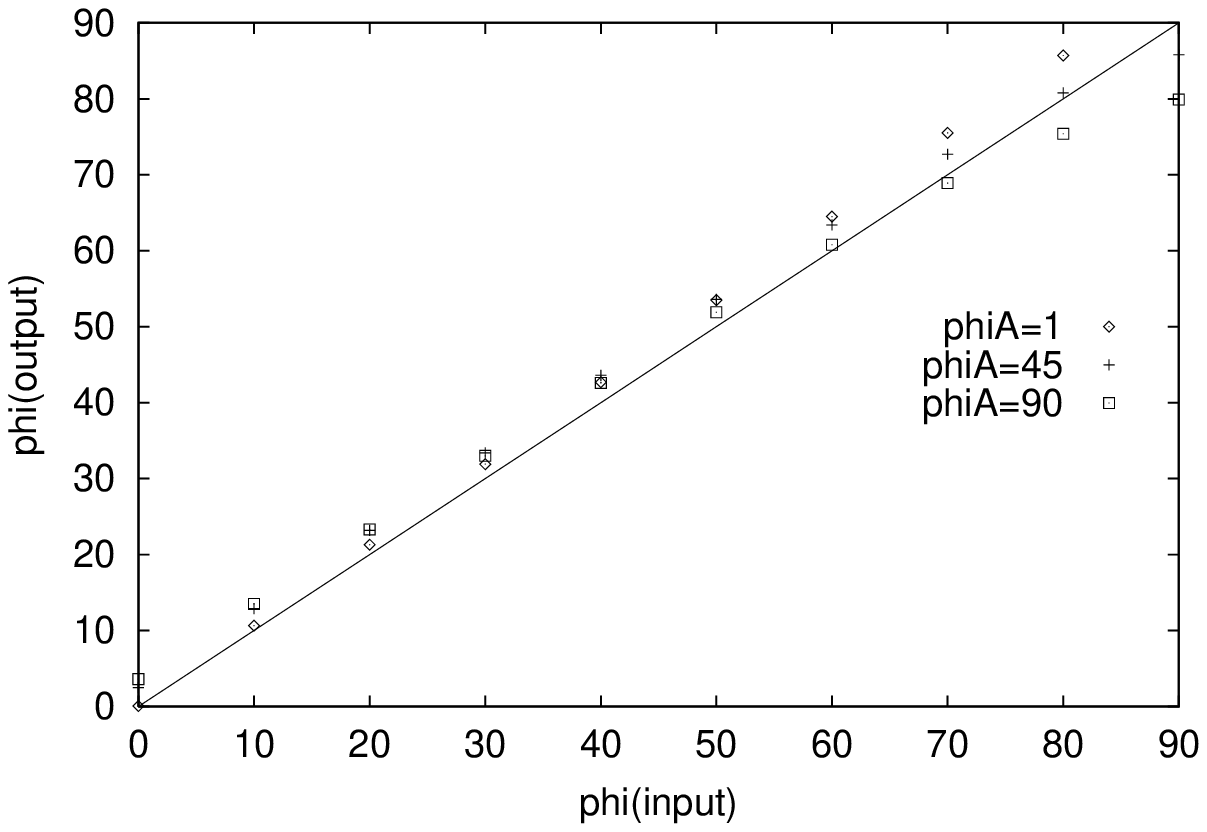,width=7cm}}\\
\hspace*{5mm}\mbox{(a)}\hspace*{7cm}\mbox{(b)}
 \end{center}
 \caption{(a) - the weak phase $\gamma$ extracted from the method using
the parameters $(R_*,\tilde A)$, as a function of the strong phase 
$\phi$ for several values of $\phi_A$ ($\epsilon_A=0.1$). 
The horizontal line shows the assumed physical value of $\gamma=76^\circ$.
(b) - the strong phase $\phi$ can be reconstructed using the $(R_*,\tilde A)$ data.}
\label{fig3}
\end{figure}

\begin{figure}[hhh]
 \begin{center}
 \mbox{\epsfig{file=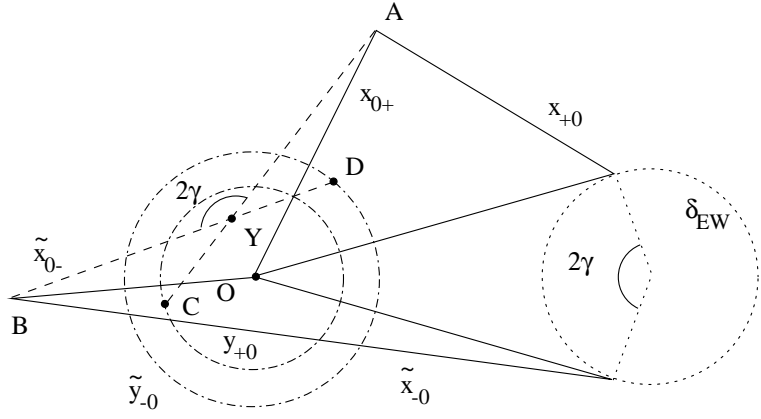,width=10cm}}
 \end{center}
 \caption{Geometric construction for the method described in Sec.~3.
$C$ and $D$ denote the intersection points of the lines $AY$ and $BY$ determined as 
explained in the text, with the two circles of radii given by $|A(B^\pm\to K^\pm K^0)|$.
}
\label{fig4}
\end{figure}

\begin{figure}[hhh]
 \begin{center}
 \mbox{\epsfig{file=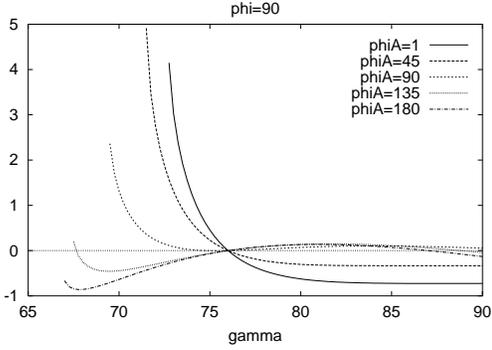,width=7cm}}
 \mbox{\epsfig{file=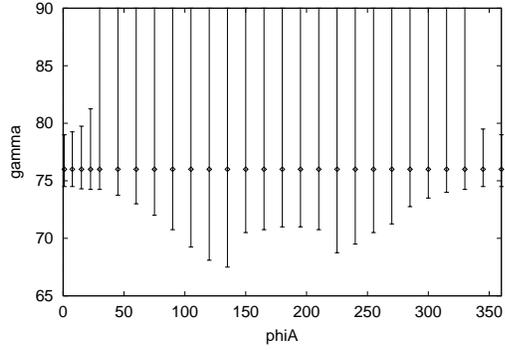,width=7cm}}\\
\hspace*{5mm}\mbox{(a)}\hspace*{7cm}\mbox{(b)}
 \end{center}
 \caption{
(a) - The left-hand of Eq.~(\ref{cond}) as a function of variable $\gamma$ for $\phi=90^\circ$
and for different values of $\phi_A$. All these curves intersect at $\gamma=76^\circ$, which
is the assumed physical value. (b) - SU(3) breaking effects introduce an error on the
extracted value of $\gamma$, here shown as function of $\phi_A$ at $\phi=90^\circ$.} 
\label{fig5}
\end{figure}

\begin{figure}[hhh]
 \begin{center}
 \mbox{\epsfig{file=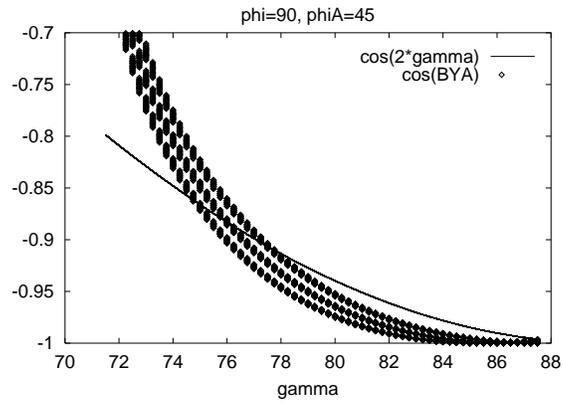,width=8cm}}
 \end{center}
 \caption{ Numerical results for the method of Sec.~4.
The two sides of the equation $\cos(BYA) = 2\gamma$ as function of
variable $\gamma$, including 30\% SU(3) breaking effects in the $p$ and
$a$ amplitudes. The physical
value of $\gamma$ is determined by the intersection of the solid line
with the wide band. The strong phases are taken as $(\phi,\phi_A)=(90^\circ,
45^\circ)$.
}
\label{fig6}
\end{figure}

\end{document}